\documentclass[%
 reprint,
 amsmath,amssymb,
prb,
]{revtex4-1}

\usepackage{graphicx}% Include figure files
\usepackage{dcolumn}% Align table columns on decimal point
\usepackage{bm}% bold math
\begin{document}

\preprint{APS/123-QED}

\title{Bragg projection ptychography of niobium phase domains}% Force line breaks with \\
%\thanks{}%

\author{Nicolas Burdet\dag}

\email{nicolas.burdet.10@ucl.ac.uk}
\affiliation{London Centre for Nanotechnology, University College London, London WC1H 0AH, UK}%Lines break automatically or can be forced with \\

\author{Xiaowen Shi\dag}%
 \affiliation{%
 Advanced Light Source, Lawrence Berkeley National Laboratory, 1 Cyclotron Rd, Berkeley, CA 94720, USA 
}%
\affiliation{Department of Physics, University of Oregon, Eugene, Oregon 97401, USA}

\author{Jesse N. Clark}%
 \affiliation{Stanford PULSE Institute, SLAC National Accelerator Laboratory, Menlo Park, CA 94025, USA}
\affiliation{Center for Free Electron Laser Science, DESY, Notkestrasse 85, Hamburg 22607, Germany}%
 \author{Xiaojing Huang}%
 \affiliation{National Synchrotron Light Source II, Brookhaven National Laboratory, Upton, New York 11973, USA}
  \author{Ross Harder}%
 \affiliation{Advanced Photon Source, Argonne, Illinois 60439, USA}
 \author{Ian Robinson}%
 \affiliation{London Centre for Nanotechnology, University College London, London WC1H 0AH, UK}
% \affiliation{Research Complex at Harwell, Harwell Campus, Oxford, OX11 0QF, UK}
 \affiliation{Brookhaven National Laboratory, Upton NY 11973, USA}
 
\thanks{Nicolas Burdet and Xiaowen Shi contributed equally to this work}


%\affiliation{
%\affiliation{
 %Third institution, the second for Charlie Author
%\affiliation{%


\date{\today}% It is always \today, today,

\begin{abstract}

Bragg projection ptychography (BPP) is a coherent x-ray diffraction imaging technique which combines the strengths of scanning microscopy with the phase contrast of X-ray ptychography.  Here we apply it for high resolution imaging of the phase-shifted crystalline domains associated with epitaxial growth.  The advantages of BPP are that the spatial extent of the sample is arbitrary, it is non-destructive and it gives potentially diffraction limited spatial resolution.  Here we demonstrate the application of BPP for revealing the domain structure caused by epitaxial misfit in a nanostructured metallic thin film. Experimental coherent diffraction data were collected from a niobium thin film, epitaxially grown on a sapphire substrate as the beam was scanned across the sample.  The data were analysed by BPP using a carefully selected combination of refinement procedures. The resulting image shows a close packed array of epitaxial domains, shifted with respect to each other due to misfit between the film and its substrate.

\end{abstract}

\pacs{Valid PACS appear here}% PACS, the Physics and Astronomy
\maketitle

%\tableofcontents

\section{\label{sec:level1}Introduction}
X-ray diffraction has long been used to investigate the properties of materials such as crystalline thin films. X-rays have the advantage over more surface-sensitive imaging modalities, such as atomic force microscopy, electron microscopy or photoemission electron microscopy because they can penetrate the entire sample.  Bragg coherent imaging methods have the Òdark fieldÓ advantage that they only consider signals from the parts of the sample that are contributing to the Bragg peak; all other sources of scattering and contributions from other components of the sample are suppressed.  The only complication in the experiment reported here is that there is a substrate Bragg peak close to the range of the data, which has to be manually removed from the data.

Synchrotron generated x-rays can be tuned to the absorption edges of specific electronic states involved in specific ordering processes. Determination of electronic, magnetic or crystalline structures are the main focus of synchrotron X-rays studies which make use of this tunability, alongside the accessibility of phase contrast, in both real and reciprocal-space imaging methods. Specifically, X-ray Coherent Diffraction Imaging (CDI) \cite{Miao99, Chapman06} is a versatile probe of nanoscale structure in non-crystalline and crystalline materials to resolutions of better than 20 nm at third-generation synchrotron facilities. The best spatial resolution achieved to date is 2 nm \cite{Takahashi10} using hard x-rays and 5 nm \cite{Shapiro14} using soft x-rays, while efforts are under way to reach sub-nanometer resolution. 

CDI uses real and reciprocal-space constraints to retrieve complex density images of nanoscale to mesoscale objects, avoiding the need for x-ray lenses.  When CDI is used on Bragg diffraction geometry, the phases of the complex image provide useful maps of strains present in the crystalline samples studied \cite{Robinson01}. To produce a complex density map of an extended sample, the beam is raster scanned across the sample with partially overlapping probe positions,  in a method called ptychography\cite{Thib, Rodenburg}. This method is found to be sufficiently robust to allow full deconvolution of the detailed features of an object from those of the probe along with any positional alignment uncertainties \cite{Guizar08, Shenfield11, Maiden12, Zhang13} and decomposition of illumination modes in case of a reduced degree of coherence \cite{Thibault13}. Ptychography, coupled with tomographic techniques, has the added advantage of measuring the 3D distribution without strenuous sample preparation.  Ptychography in the Bragg-geometry is able to discriminate between areas in the sample with different ordering strengths or orientations, such as domain structures in thin films.  

\begin{figure*}
\begin{center}
\includegraphics[width = 16 cm]{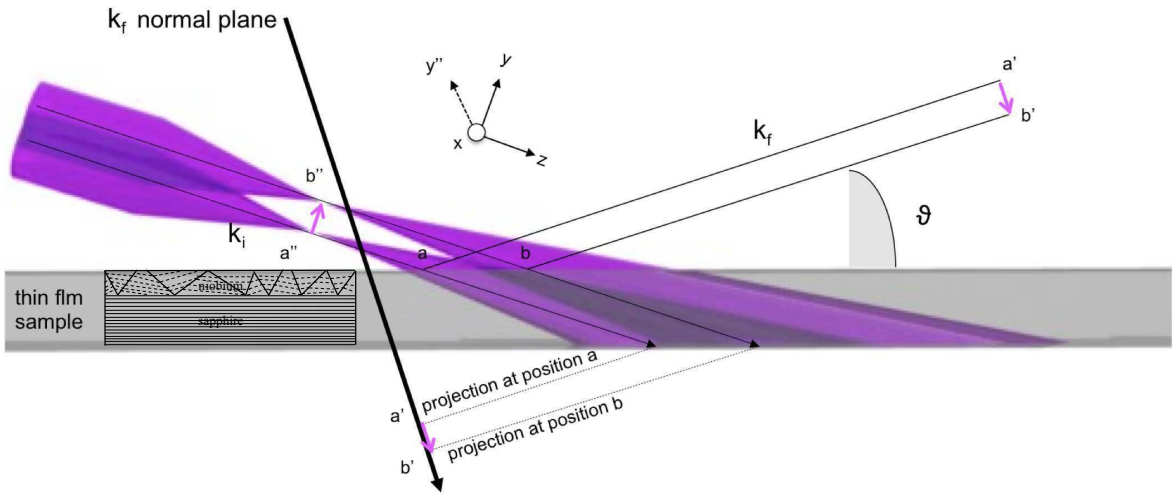}% Here is how to import EPS art
\caption{\label{fig:1} Schematic of Bragg projection on an idealised niobium thin film presenting a single layer of structural domains. By scanning the beam in the laboratory frame, the projected overlap after the beam displacement is symmetric with equal step-sizes within the detector plane ($|a'b'| = -|a''b''|$) for the specular case, as used in the work of Hruszkewycz et. al. \cite{Hruszkewycz12}}
\end{center}
\end{figure*}

Nonetheless, the realization of Bragg Projection Ptychography (BPP) by augmenting ptychography to Bragg-geometry experiments is non-trivial due to the complexities of the experimental geometries involved. The technique is still under development because of its great potential towards gaining high-resolution images over wide fields-of-view. The recent development of BPP by Hruszkewycz et. al. \cite{Hruszkewycz17} incorporates the geometrical relation between the Bragg measurements and the overlapping projections by inverting only a set of 2D diffraction patterns (a single cut through the 3D diffraction volume) and performing judicious scans. In our realization of BPP, we use a set of 3D piezo-stages to perform a laboratory frame scan that preserves the perpendicularity to the probe directed along the  optical axis, $\mathbf{k_i} \perp$, independently of the Bragg angle. In this manner, the projected displacements of the sample are symmetric and of equal magnitudes on the detector plane, normal to $\mathbf{k_f}$, as seen by $-y''=y$ in Fig.~\ref{fig:1}. 

The structure of the thin film sample is assumed to contain a single layer of domains within the penetration depth which is projected along the propagation direction. Both conditions are of critical importance in order to avoid integration of multiple domains along the $\mathbf{k_f}$ direction and so to keep the relationship between the reconstructed Bragg projection and sample structure satisfied, as recently considered in the work of Hruszkewycz et. al.\cite{Hruszkewycz16}.  BPP has been shown to be a sensitive tool for measurements and characterizations of lattice distortions in thin films \cite{Holt14, Hruszkewycz12, Hruszkewycz13, ChamardSci2015, Pateras2015, Godard2011}, with recent development to three dimensions (3D)\cite{Hruszkewycz17}. Earlier investigations on the structure of thin films with coherent X-ray diffraction methods failed to be conclusive \cite{Stadler07, BeanTh}, so we present our progress here to investigate the application of BPP to the problem of phase domain structures in thin films. The Nb thin film was grown by K. Ritley in the laboratory of C. P. Flynn at the university of Illinois, Urbana-Champaign, and the thickness of the Nb thin film was measured by an optical profilometer to be 100 nm. The beam size is about 800 nm and it is around 3 times bigger on the sample since it is a reflection geometry experiment. The ratio of beam size on sample to the thin film thickness is about 25, which means our geometry is mathematically valid for Bragg projection ptychography. 

Both of our work and Hruszkewycz et. al. \cite{Hruszkewycz17} have Strong divergent beam, however, in our work, we have only a section of the divergent beam. We have used section projection theorem. We cut center of speckle, Fourier transformed into a projection view of domains. The difference between our work and the work of \cite{Hruszkewycz17}: In our work, we are not modelling the probe and just image the probe, and our probes change more and our probe fluctuation is more intense, where in \cite{Hruszkewycz17}, their probe is more stable. There are several major distinctions.  Our experiment looks at a granular domain structure in a thin film, whereas \cite{Hruszkewycz17} look at an almost perfect crystalline film.  Our beam divergence is much smaller than that of \cite{Hruszkewycz17}, so that the diffraction pattern of our samples extends far outside the beam divergence, wheras that of \cite{Hruszkewycz17} stays mostly within the beam. Since the two experiments are in different limits, they benefit from different approximations.

\section{\label{sec:level2}Niobium thin films}

Niobium films have been found numerous applications thanks to their ability to grow on sapphire substrates of various surface orientations, providing valuable buffer layers for the synthesis of layered structures in many areas of nanotechnology \cite{Wildes01, Nb2012}. Niobium can be epitaxially grown on a variety of ceramic materials, MgO, GaAs, InAs by means of molecular beam epitaxy but so far it is the Nb/Sapphire system that produces the highest crystalline quality. (11-20) oriented sapphire substrates are closely matched to the spacing of body-centred cubic (BCC) niobium (110) planes. The sample used in this experiment was a Nb(110) thin film with a thickness of about 100 nm, which is above the critical thickness \cite{IKR03,Barabash01}. Under these conditions, the real nanostructure of these systems is generally complicated by the unavoidable lattice misfit between the film and substrate, which gives rise to elastic strain that is relaxed by misfit dislocations \cite{Flynn88}. Such epitaxial structures have already been investigated by X-ray diffraction, however so far without the ability to image the individual domains\cite{Wildes01}. BPP is an effective way to understand such structures at the level of the nanoscale domains, because the inherent real-space phase-contrast images of the thin film can identify these domains as crystal blocks with distinct phases. This allows comparison with models of domain structures where the global elastic strain of the film-substrate system is minimized. 

%\subsubsection{Wide text (A level-3 head)}
%The \texttt{widetext} environment will make the text the width of the
%full page, as on page~\pageref{eq:wideeq}. (Note the use the
%\verb+\pageref{#1}+ command to refer to the page number.) 
%\paragraph{Note (Fourth-level head is run in)}
%The width-changing commands only take effect in two-column formatting. 
%There is no effect if text is in a single column.

\section{\label{sec:level3}Experiments}
\begin{figure}[b]
\includegraphics[width = 8.6 cm]{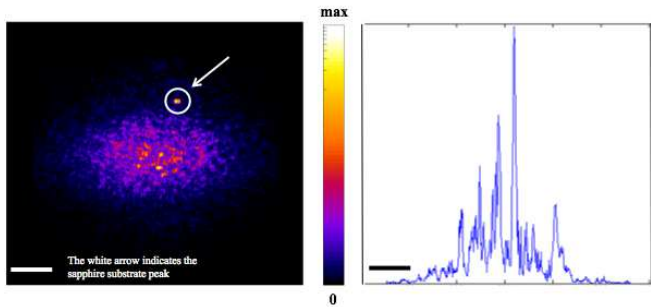}% Here is how to import EPS art
\caption{\label{fig:2} Left: logarithmic plot of a typical coherent diffraction pattern taken at the Nb(110) specular Bragg peak, with maximum spatial frequency of $ \sim33$ $\mu$ $m^{-1}$. The tail of the diffraction peak of the sapphire substrate is indicated with a circle and arrow. Right: cross sectional profile taken along the detector $x$ direction at the centre of the peak. The speckle features are of high contrast with a high degree of coherence and have $\sim12$ pixels per speckle. Scale bar, $ \sim10$ $\mu$ $m^{-1}$}
\end{figure}

Coherent diffraction patterns from a specular (110) reflection were collected with focused 8.9 keV X-rays, in a concentric scan pattern, using a Charge-coupled Device (CCD) X-ray detector positioned 2.184 m downstream of the sample. A highly coherent beam was generated with a $100\mu m$ horizontal secondary-source aperture, $26m$ from the undulator source.  This was focused to $\sim 800\times800$ nm$^2$ by a pair of Kirkpatrick-Baez (KB) mirrors with entrance slits set to $30_{\text{H} } \times 50_{\text{V} } $ $\mu$m$^2$. The specular Nb(110) Bragg reflection was chosen for ptychographic scan. The crystalline structure of niobium is body centered cubic lattice with a lattice parameter of 0.3301 nm, and Sapphire has unit cell length a = 0.350 nm. The lattice mismatch of these in reciprocal-space was calculated by equation (1) of Ref \cite{Xiaowen2012} with the experimental parameters used in the experiment such as x-ray energy, sample to detector distance and detector pixel size etc. This is roughly in agreement with the experimental coherent diffraction data shown in (Fig.~\ref{fig:2}\.to be around $0.1 $nm$^{-1}$. The XRD result in Fig. 3 in Ref \cite{Nb2012} suggesting the separation of Sapphire bulk peak and the Nb thin film Bragg peak was around $0.1- 0.2 $nm$^{-1}$, which is in agreement with our experimental diffraction pattern displayed in (Fig.~\ref{fig:2}\. of our paper. We would like to note that the XRD result in Fig 3 in Ref \cite{Nb2012} was performed that was solely sensitive in the thickness direction of the Nb thin film, whereas our measurements mainly contain x-ray diffraction patterns in the Nb thin film surface direction, not the Nb thin film thickness direction as measurements with conventional XRD methods, thus the results from the two techniques can only be compared with some error-bars allowed in reciprocal space. 

A typical diffraction pattern is shown in Fig.~\ref{fig:2}. and a summary of experimental parameters is given in Table.~\ref{tab:table1}.

\begin{table}
\caption{\label{tab:table1}
Experimental parameters for ptychographic scans on Nb(110) data taken at APS beamline 34-ID-C.}

\begin{tabular}{lll}
  Properties & Value  & unit \\
  \hline
  Energy & 8.9 & KeV \\
  $\lambda$ & $1.39$ & \AA \\
  lattice parameter, a & 3.3 & \AA \\
  sample at $ \theta_{110}$  & 17 & $^{\circ}$ \\
  camera at $ \delta_{110}$ & 34.7 & $^{\circ}$ \\
  beam size & $0.8_H$ $ \times$ $0.8_V$ & $\mu$m \\
  detector distance, $z$ & 2.184 & m \\
  detector pixel size, $\Delta_p $ & 20 & $\mu$m \\
  pixel numbers  & $512 \times 512$ & pixels \\
  real-space pixel & $30$ & nm \\
  scan range & $2 \times 2$ & $\mu$m \\
  scan step size & $150$ & nm \\
  degree of overlap  & $80$ & \% \\
  \hline
\end{tabular}

\end{table}

Mutual interference between different regions of the film, when coherently illuminated, modulates the large Nb peak into a densely speckled pattern with the mean feature size inversely proportional to the incident beam size. The beam also diffracts from crystalline substrate and gives rise to a sharp single Bragg peak which is of comparable intensity to the central speckles arising from the thin film peak.  This substrate peak was removed from the data by manually setting the intensity to zero. The visibility of the speckle was estimated of $80 \pm 5\%$ from a line plot (Fig.~\ref{fig:2}\ showing good contrast. The speckles are oversampled with approximately $\sim12$ pixels across each feature.  The mismatch between the bulk peak and the Nb speckle patterns, which gives rise to the creation of phase domains in the system, also causes the peak separation in reciprocal-space seen on the CCD detector, so that we were able to isolate the Nb thin-film speckle patterns for ptychographic reconstructions. From our coherent diffraction pattern displayed in Fig. 2, we calculate that the average domain size should have 3 to 1 ratio with the larger side of around average size of 100 nm. The average size of long side of phase domains of Nb thin film of 100 nm with long to short size of being 3 to 1 is in agreement with the Atomic Force Microscopy (AFM) image reported in Fig. 1(b) of Ref \cite{Nb2012} with the same Nb thin film thickness.

\section{\label{sec:level4}Ptychographic analysis}
The first step of the ptychographic reconstruction was to estimate the complex probe illumination function by reconstructing a classical Siemens star test pattern in the transmission geometry.  The test pattern was a strongly phase-shifting radial-spoke design cut into a 1.5-micron thick W film on a $Si_3N_4$ membrane using e-beam lithography.  After 60 iterations of the difference-map (DM) algorithm\cite{Thib}, an additional step was applied to update the probe function for 300 more iterations. The first reconstruction attempt was found to be clean, without distortions arising from the objectÕs structural features, yet it showed rather weak unresolved amplitudes away from the central region.  Another undesirable misbehavior of the algorithm led to a filling in of the corners of the probe array.  To avoid this, only a circular region in the center of the array was retained after the probe update step, acting as a ``support" function.  As drift and uncertainties in the probe positions is known to occur in ptychography reconstructions, an additional 100 iterations of probe positions corrections \cite{Maiden12} was attempted following the previous ptychographic reconstructing scheme. However, no improvement in the probe was noticed, and the positions were found to drift far from their origin. The mean error in scan positions was calculated with:

\begin{equation}
\Delta r_{ \text{avg} }^{  \text{ final }  } = \frac{ \sum_j | \mathbf{r}_{j}^{ \text{ final } } -  \mathbf{r}_{j}^{ \text{scan} }  | }{N} = \frac{ \sum_j | \mathbf{c}_{j}^{ \text{final}} | }{ N } , 
\end{equation}

where $N$ is the number of positions and $ \mathbf{c}_{j}$ the deviation from the idealized scan. 

One of the fundamental requirements for ptychography to reliably factorize the real space image plane into a probe function and an object function is that the probe structure has to be same at each position\cite{Rodenburg04}.  This requires a stable experimental setup so that the probe is time-invariant.  In reality, shape and intensity variations in the probe structure were found to occur for scans approaching an hour in duration. It was considered possible, therefore, that the poorly formed outer fringes found in the first attempt to reconstruct the probe could have been due to probe instabilities. To attempt to correct for this effect, the DM real-space update equations (see Eq.7-8 in \cite{Thib}) were modified to include a sequence of successive probe functions  $ P (\mathbf{r} ) \rightarrow P_{d}(  \mathbf{r} )$ during reconstruction of a single object $O(  \mathbf{r} )$ with: 
\begin{equation}
P_{d}(  \mathbf{r} ) = \frac{\sum_{d=1}^{D} \sum_{j = (d-1)N/D+1}^{(d)N/D} O^*( \mathbf{r} - \mathbf{r_j} ).\psi_d( \mathbf{r}  )}{\sum_{d=1}^{D} \sum_{j = (d-1)N/D+1}^{(d)N/D} | O(  \mathbf{r} -  \mathbf{r_j} )|^2} , 
\end{equation} 

and

\begin{equation} \label{dynamique}
O(  \mathbf{r} ) = \frac{\sum_{d=1}^{D} \sum_{j = (d-1)\frac{N}{D}+1}^{d\frac{N}{D}+1} P_{d}^*( \mathbf{r}+  \mathbf{r_j} ). \psi_d( \mathbf{r} )}{\sum_{d=1}^{D} \sum_{j = (d-1)\frac{N}{D}+1}^{d\frac{N}{D}+1} |P_{d}^*( \mathbf{r}+  \mathbf{r_j} )|^2} , 
\end{equation}

where $D$ is the number of illuminations to be reconstructed and $N$ the number of scan points.   

\begin{figure*}
\begin{center}
\includegraphics[width = 12 cm]{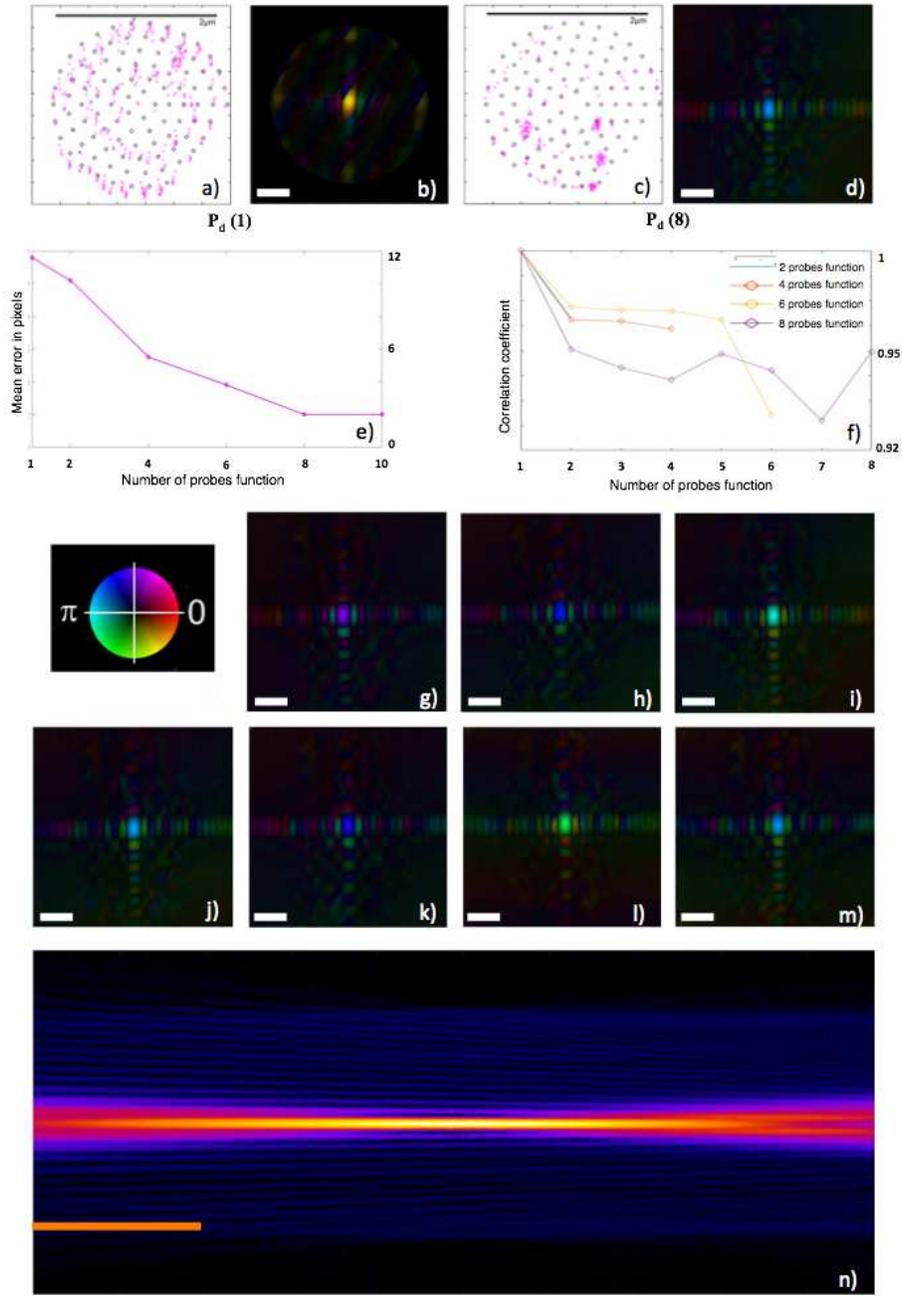}% Here is how to import EPS art
\caption{\label{fig:4} a) plot of the corrected positions for the reconstruction of Nb(110) data with a single probe function ($D = 1$). The points are plotted in violet are shown every 5 iterations during the refinement process.  b) reconstructed probe for $D=1$, scale bar, $2\mu m$. c) plot of the corrected positions for $D = 8$ probes.  d) first probe function for $D=8$, scale bar, $2\mu m$. e) plot of the mean error ($\Delta_{avg}^{final}$) as a function of $D$.  f) correlation coefficient plot of the probe functions for reconstructions with $D = 2,4,6$ and $8$ respectively. g)-m) the remaining $7$ probes for the $D = 8$ reconstruction. n) the propagation of KB focused probe in this study. The yellow scale bar corresponds to 2 mm distance. }
\end{center}
\end{figure*}

While this method bears some similarity with the parallelization strategy proposed by Guizar-Sicairos et. al. \cite{Nashed14, Guizar14}, although it involves no synchronization and stitching processes to obtain bigger field-of-view in 2D to avoid probe fluctuations. Therefore, it is possible to make self-consistent local data, so that probe does not fluctuate during data acquisition during one single ptychographic scan. There is another recent work focusing on this probe fluctuation topic \cite{Odstrcil16}, In this paper, the authors used coherent probes from single value decomposition of original probe, and this recently developed method is probably a more elegant way to address the issue of fluctuating probe, and the method should be used in the future. This technique will generate fewer ambiguities in the reconstructed phase domains.   It also should also not be confused with the multimodal decomposition method\cite{Thibault13} which is needed in the case of a partially coherent illuminating probe. The multi-modal description of the beam was not found to be necessary in our experiments, although it could be implemented in the future. Since the concentric scan pattern was decomposed in a linear fashion, it resulted in an overlap constrained only in one lateral direction at the outside of the spiral scan.  This multi-probe method was tested by performing reconstructions with sequences of 2, 4, 6 and 8 independent probe functions. The reconstructions followed the following algorithm sequence: for the first 60 iterations, the reconstructions were performed using the original probe function estimate from the test sample, which was updated for 40 iterations before starting the dynamical probe scheme to evaluate $D-1$ further probe functions.  This scheme was run for 300 iterations and followed by 100 iterations of position corrections. The carefully selected combination of refinement that includes multiple probes analysis and position correction. The position correction method we used is the annealing method \cite{MaidenA2012}, because this method seems to be robust to this case, in conjunction with multiple probes analysis. Each of the ptychography scan we performed is around one hour, which suggests that for every 1/8 hour the probe has changed. That means we had probe fluctuations for every 7.7 mins on average. Reconstructions performed with $D=2$ (two independent probe functions) yielded noticeable improvement with fringes that were more extended and regular.  However, it was found that $D=8$ led to probe functions with the clear $\frac{\sin(x)}{x}$ shape expected for a pair of KB mirrors at its focal plane.  These images are shown in Fig.~\ref{fig:4}\.\ \cite{BurdetSci2016}

The plot reported in Fig.~\ref{fig:4}\.\ (e) clearly shows the mean error in the determined probe positions decreases as a function of $D$ with a minimum of 2 pixels (60 nm) reached for the $D=8$ case, for which it can be seen that the positions migrate in an unbiased way around the starting positions with no apparent global trend. The $x$-direction was found to be more stable than $y$, which may be coupled to the observation that the probe fringes are better defined in this direction. The first probe ($d=1$) of the $D=8$ probe function is displayed in Fig.~\ref{fig:4}\.\ (d). Thus the position correction seems to have little influence on the probe functions, which is also clear from our finding for $D=1$ in Fig.~\ref{fig:4}\.\ (b).  The position deviation from one probe function to the next are within 5 \% according to the correlation plot in Fig.~\ref{fig:4}\.\ (f), and the total drift was less than 8 \% over the full duration the ptychographic measurement.  It is believed that this refinement scheme does not increase the number of degrees of freedom in the ptychographic algorithm but rather provides a better set of constraints to drive convergence.  Overall the reproducibility of the result was better with 8 probe functions reconstructions involving a smaller number of probe functions. The 8 probe functions are images of probes at different points in time, which is the average probe of this section of the scan area.   

In the final images of the sample, the removal of linear phase offsets is usually not found to be an issue, since this is unconstrained.  However, different phase offsets that arose from different random seeding of the algorithm did introduce possible ambiguities into the final reconstruction. The various phase domain images of the sample were therefore not unique because they contained an arbitrary phase offset. Guizar-Sicairos et. al. \cite{Guizar11} have addressed the fact that an additional ambiguity arises in the form of linear phase ramps whenever both object and probe are reconstructed: however, to satisfy both the overlap constraint and Fourier modulus constraint, these linear phase ramps must in principle be of equal magnitude and opposite sign. Hence, phase ramps found in each of the $D$ reconstructed illuminations were calculated by the center of mass method and used to cancel out phase ramps present within the complex object. Their respective values ($\phi_d$) are reported in Table.~\ref{tab:table2}\.~for the ptychographic reconstructions using a subset of $8$ probe functions.

\begin{table}
\caption{\label{tab:table2}
Phase values of ramps found in two random seeding of ptychographic reconstructions using a subset of 8 probe functions respectively.   }
\begin{tabular}{ccccccccc}
  $P_d$ & $P_1$ & $P_2$ & $P_3$ & $P_4$ & $P_5$ & $P_6$ & $P_7$ & $P_8$ \\
  \hline
   $\phi_d$  & $0.2\pi$ & $0.1\pi $ & $0.2\pi$  & $0\pi$  & $-0.2\pi$  & $0.3\pi$  & $-0.3\pi$ & $-0.1\pi$  \\
   $\phi_d$  & $- 0.5\pi$ & $- 0.4\pi $ & $0.2\pi$  & $-0.4\pi$  & $0.2\pi$  & $0.1\pi$  & $0.4\pi$ & $0.1\pi$  \\
  \hline
\end{tabular}
\label{illuminees}
\end{table}

As expected, the phase ramp values were found to change from run to run but also in between each probe function of the same iteration and reconstruction. A change between neighboring illumination phase ramps further complicates the phase ramp removal. Nonetheless, an integral phase ramp cancellation on each of the $j=N$ views object's subfields could be performed according to:

\begin{equation} \label{suppression}
O_{cc}( \mathbf{r}) =  \sum_{d=1}^{D} \sum_{j = (d-1)\frac{N}{D}+1}^{d\frac{N}{D}}  e^{- \phi_d} . O(  \mathbf{r}-  \mathbf{r_j} +  \mathbf{c}_{j}^{ \text{final} }) , 
\end{equation} 

where $\phi_d$ is the associated phase ramp array. 

\begin{figure}[b]
\includegraphics[width = 8.5 cm]{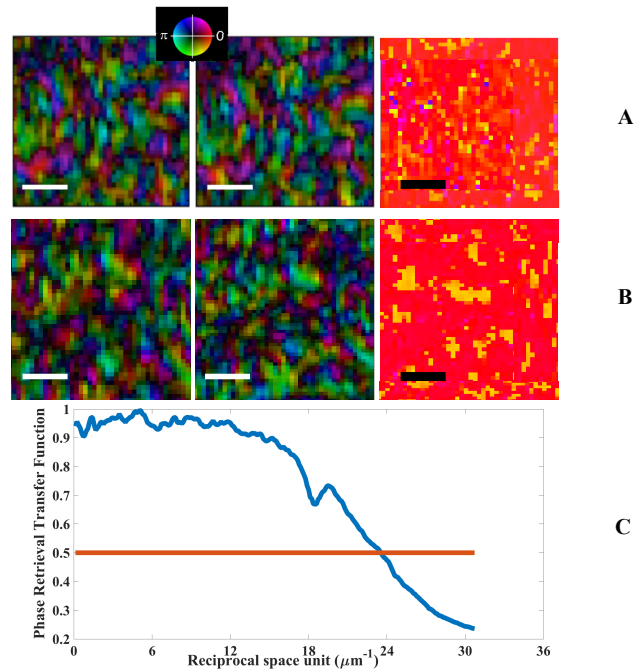}% Here is how to import EPS art
\caption{\label{fig:5} : A: Hue rendering of central region of objects with suppressed phase ramps, according to Eq.\ref{suppression} and phase difference map of the two reconstructions with random initial starts. The difference map image was performed on the image shown. B: same for a different sample area. Both white and black scale bars are 200 nm. C: Phase Retrieval Transfer Function of the data used in this study. }
\end{figure}

The final phase image of the sample, at two different regions near the center, are displayed in Fig.~\ref{fig:5}\. We leave the reconstructions in the xyÕÕ, which is the detector frame. To demonstrate the reliability of measured and calculated intensity patterns, we have calculated the Phase Retrieval Transfer Function (PRTF) in Fig.~\ref{fig:5}\. C. PRTF was calculated \cite{Shapiro05} by comparing the measured and calculated diffraction patterns at the same reciprocal space Q-region, and the ratio of two intensities was analysed to be between 0 and 1, with 1 being the most faithful match between the two intensities. This allows us to estimate the resolution of the final image to be $37nm$. The reconstructed images are displayed in the $xyÕÕ$ coordinate frame defined in Fig.~\ref{fig:1}\, which is not the coordinate frame of the sample face. The sample features are therefore elongated due to the footprint of x-ray beam in the specular reflection geometry used in the experiment. Before applying the subtraction of two phase images, the images are sub-pixel shifted by maximizing the modulus square of the cross-correlation of the two images in this study as illustrated in the \cite{subpixel}. The complex images are registered before calculation of difference map. Difference map is calculated as: Phase of image 1 - Phase of image 2. We note that there are small, apparently random, differences between the reconstructions from multiple random starts, as presented in Fig.~\ref{fig:5}\., we attribute these to differences in the propagation of noise from the original data.   

The image of Fig.~\ref{fig:5}\. can be readily understood as a phase-contrast picture of the nanoscale domains in the Nb(110) thin film studied by BPP.  It shows a mosaic of nanoscale blocks of material, shifted in phase with respect to each other.  Within each mosaic block, the phase is roughly constant, indicating it is a rather perfect, unstrained piece of crystal.  The boundaries between the blocks are abrupt and quite straight, with a slight dimming of the amplitude that is understood as partial cancelation of different phases falling within a resolution element. These boundaries can be understood as the slip planes between the crystal blocks where the misfit dislocations are accumulated (but not resolved).  It can be seen that the blocks have a fairly regular size and a spacing of about 40nm.  The misfit between the Nb and the Sapphire substrate should determine the dislocation density, but this has to be considered at the high growth temperature where diffusion is activated.  The amount of misfit can be estimated from the observation that the Nb spacing slips by $a_0 / \sqrt{2} = 0.23nm$ between every block, spaced 40nm apart.  This corresponds to a misfit of $0.6\%$.  The AFM image reported in Fig. 1(b) in Ref \cite{Nb2012}, displaying the average size of long side of phase domains of Nb thin film of 100 nm with long to short size of being 3 to 1 with the same Nb thin film thickness shows their result is in agreement with our coherent x-ray scattering intensities and BPP reconstructions. We note that while these phase domain structures are expected from theories of epitaxy, the misfit structures and are rather difficult to detect with other methods: for example, the aggressive sectioning of the sample, required for TEM, can disturb the delicate strained structures in the thin film. We envisage in the future, we will combine coherent probe decomposition \cite{Odstrcil16} with 3D Bragg Projection Ptychography to help better understanding of structures and rotations of reconstructed phase domains. Our independent probes have their own phase ramps, this technique could remove the probe phase ramps, so that one can better understand the variations of phase structures in the reconstructed sample without phase ramps.  

\section{\label{sec:level5}Summary and conclusions}
We have demonstrated the application of BPP in reflection geometry on a sample that contains crystallographic  domains which appear as phase-shifted blocks in the reconstructed image at a resolution of $37nm$.  Direct imaging of such domains, expected from the presence of misfit structures in epitaxial thin film heterostructures, through the phase contrast channel is a new result.  BBP has the potential for imaging a wide variety of crystalline thin films of scientific interest, with both tens-of-nanometers resolution and an arbitrarily wide field of view.  Given the large amount of information contained in coherent diffraction patterns, such as Fig.~\ref{fig:2}\ , and the results presented here about the complexity of the modeling needed to produce images, the full potential of BPP will call for even more advanced iterative algorithms to be developed and implemented. For example, in the case of less symmetric geometries, more understanding of the beam coherence properties is needed. Nevertheless, this work has advanced the experimental procedures and algorithm development for Bragg projection ptychography in the reflection geometry. Important future applications might include the imaging of orbital ordering domains in strongly correlated electron systems, or spin density waves, for example, once suitable cryostats can be implemented.

\begin{acknowledgments}
Use of the Advanced Photon Source, an Office of Science User Facility operated for the U.S. Department of Energy, Office of Science by Argonne National Laboratory, was supported by the U.S. DOE under Contract No. DE-AC02-06CH11357.  Beamline 34-ID-C was built with a grant from the National Science Foundation DMR-9724294. This research was carried out under grants EP/G068437/1 and EP/I022562/1 from the UK Engineering and Physical Sciences Research Council (EPSRC). X. S. acknowledge partial supported by the U.S. Department of Energy, Office of Basic Energy Sciences, Division of Materials Science and Engineering, under Grant No. DE- FG02-11ER46831. J.N.C. gratefully acknowledges financial support from the Volkswagen Foundation. Work at Brookhaven National Lab was supported by the U.S. Department of Energy, Division of Materials Science, under Contract No. DE-SC00112704.
\end{acknowledgments}

%\nocite{*}

\bibliography{apssamp}% Produces the bibliography via BibTeX.

\begin{thebibliography}{37}%
\makeatletter
\providecommand \@ifxundefined [1]{%
 \@ifx{#1\undefined}
}%
\providecommand \@ifnum [1]{%
 \ifnum #1\expandafter \@firstoftwo
 \else \expandafter \@secondoftwo
 \fi
}%
\providecommand \@ifx [1]{%
 \ifx #1\expandafter \@firstoftwo
 \else \expandafter \@secondoftwo
 \fi
}%
\providecommand \natexlab [1]{#1}%
\providecommand \enquote  [1]{``#1''}%
\providecommand \bibnamefont  [1]{#1}%
\providecommand \bibfnamefont [1]{#1}%
\providecommand \citenamefont [1]{#1}%
\providecommand \href@noop [0]{\@secondoftwo}%
\providecommand \href [0]{\begingroup \@sanitize@url \@href}%
\providecommand \@href[1]{\@@startlink{#1}\@@href}%
\providecommand \@@href[1]{\endgroup#1\@@endlink}%
\providecommand \@sanitize@url [0]{\catcode `\\12\catcode `\$12\catcode
  `\&12\catcode `\#12\catcode `\^12\catcode `\_12\catcode `\%12\relax}%
\providecommand \@@startlink[1]{}%
\providecommand \@@endlink[0]{}%
\providecommand \url  [0]{\begingroup\@sanitize@url \@url }%
\providecommand \@url [1]{\endgroup\@href {#1}{\urlprefix }}%
\providecommand \urlprefix  [0]{URL }%
\providecommand \Eprint [0]{\href }%
\providecommand \doibase [0]{http://dx.doi.org/}%
\providecommand \selectlanguage [0]{\@gobble}%
\providecommand \bibinfo  [0]{\@secondoftwo}%
\providecommand \bibfield  [0]{\@secondoftwo}%
\providecommand \translation [1]{[#1]}%
\providecommand \BibitemOpen [0]{}%
\providecommand \bibitemStop [0]{}%
\providecommand \bibitemNoStop [0]{.\EOS\space}%
\providecommand \EOS [0]{\spacefactor3000\relax}%
\providecommand \BibitemShut  [1]{\csname bibitem#1\endcsname}%
\let\auto@bib@innerbib\@empty
%</preamble>
\bibitem [{\citenamefont {Miao}\ \emph {et~al.}(1999)\citenamefont {Miao},
  \citenamefont {Charalambous}, \citenamefont {Kirz},\ and\ \citenamefont
  {Sayre}}]{Miao99}%
  \BibitemOpen
  \bibfield  {author} {\bibinfo {author} {\bibfnamefont {J.~W.}\ \bibnamefont
  {Miao}}, \bibinfo {author} {\bibfnamefont {P.}~\bibnamefont {Charalambous}},
  \bibinfo {author} {\bibfnamefont {J.}~\bibnamefont {Kirz}}, \ and\ \bibinfo
  {author} {\bibfnamefont {D.}~\bibnamefont {Sayre}},\ }\href {\doibase
  10.1038/22498} {\bibfield  {journal} {\bibinfo  {journal} {Nature}\ }\textbf
  {\bibinfo {volume} {400}},\ \bibinfo {pages} {342} (\bibinfo {year}
  {1999})}\BibitemShut {NoStop}%
\bibitem [{\citenamefont {Chapman}\ \emph {et~al.}(2006)\citenamefont
  {Chapman}, \citenamefont {Barty}, \citenamefont {Marchesini}, \citenamefont
  {Noy}, \citenamefont {Hau-Riege}, \citenamefont {Cui}, \citenamefont
  {Howells}, \citenamefont {Rosen}, \citenamefont {He}, \citenamefont {Spence},
  \citenamefont {Weierstall}, \citenamefont {Beetz}, \citenamefont {Jacobsen},\
  and\ \citenamefont {Shapiro}}]{Chapman06}%
  \BibitemOpen
  \bibfield  {author} {\bibinfo {author} {\bibfnamefont {H.~N.}\ \bibnamefont
  {Chapman}}, \bibinfo {author} {\bibfnamefont {A.}~\bibnamefont {Barty}},
  \bibinfo {author} {\bibfnamefont {S.}~\bibnamefont {Marchesini}}, \bibinfo
  {author} {\bibfnamefont {A.}~\bibnamefont {Noy}}, \bibinfo {author}
  {\bibfnamefont {S.~R.}\ \bibnamefont {Hau-Riege}}, \bibinfo {author}
  {\bibfnamefont {C.}~\bibnamefont {Cui}}, \bibinfo {author} {\bibfnamefont
  {M.~R.}\ \bibnamefont {Howells}}, \bibinfo {author} {\bibfnamefont
  {R.}~\bibnamefont {Rosen}}, \bibinfo {author} {\bibfnamefont
  {H.}~\bibnamefont {He}}, \bibinfo {author} {\bibfnamefont {J.~C.~H.}\
  \bibnamefont {Spence}}, \bibinfo {author} {\bibfnamefont {U.}~\bibnamefont
  {Weierstall}}, \bibinfo {author} {\bibfnamefont {T.}~\bibnamefont {Beetz}},
  \bibinfo {author} {\bibfnamefont {C.}~\bibnamefont {Jacobsen}}, \ and\
  \bibinfo {author} {\bibfnamefont {D.}~\bibnamefont {Shapiro}},\ }\href
  {\doibase 10.1364/josaa.23.001179} {\bibfield  {journal} {\bibinfo  {journal}
  {Journal of the Optical Society of America a-Optics Image Science and
  Vision}\ }\textbf {\bibinfo {volume} {23}},\ \bibinfo {pages} {1179}
  (\bibinfo {year} {2006})}\BibitemShut {NoStop}%
\bibitem [{\citenamefont {Takahashi}\ \emph {et~al.}(2010)\citenamefont
  {Takahashi}, \citenamefont {Nishino}, \citenamefont {Tsutsumi}, \citenamefont
  {Zettsu}, \citenamefont {Matsubara}, \citenamefont {Yamauchi},\ and\
  \citenamefont {Ishikawa}}]{Takahashi10}%
  \BibitemOpen
  \bibfield  {author} {\bibinfo {author} {\bibfnamefont {Y.}~\bibnamefont
  {Takahashi}}, \bibinfo {author} {\bibfnamefont {Y.}~\bibnamefont {Nishino}},
  \bibinfo {author} {\bibfnamefont {R.}~\bibnamefont {Tsutsumi}}, \bibinfo
  {author} {\bibfnamefont {N.}~\bibnamefont {Zettsu}}, \bibinfo {author}
  {\bibfnamefont {E.}~\bibnamefont {Matsubara}}, \bibinfo {author}
  {\bibfnamefont {K.}~\bibnamefont {Yamauchi}}, \ and\ \bibinfo {author}
  {\bibfnamefont {T.}~\bibnamefont {Ishikawa}},\ }\href {\doibase
  10.1103/PhysRevB.82.214102} {\bibfield  {journal} {\bibinfo  {journal}
  {Physical Review B}\ }\textbf {\bibinfo {volume} {82}} (\bibinfo {year}
  {2010}),\ 10.1103/PhysRevB.82.214102}\BibitemShut {NoStop}%
\bibitem [{\citenamefont {Shapiro}\ \emph {et~al.}(2014)\citenamefont
  {Shapiro}, \citenamefont {Yu}, \citenamefont {Tyliszczak}, \citenamefont
  {Cabana}, \citenamefont {Celestre}, \citenamefont {Chao}, \citenamefont
  {Kaznatcheev}, \citenamefont {David}, \citenamefont {Maia}, \citenamefont
  {Marchesini}, \citenamefont {Meng}, \citenamefont {Warwick}, \citenamefont
  {Yang},\ and\ \citenamefont {Padmore}}]{Shapiro14}%
  \BibitemOpen
  \bibfield  {author} {\bibinfo {author} {\bibfnamefont {D.~A.}\ \bibnamefont
  {Shapiro}}, \bibinfo {author} {\bibfnamefont {Y.-S.}\ \bibnamefont {Yu}},
  \bibinfo {author} {\bibfnamefont {T.}~\bibnamefont {Tyliszczak}}, \bibinfo
  {author} {\bibfnamefont {J.}~\bibnamefont {Cabana}}, \bibinfo {author}
  {\bibfnamefont {R.}~\bibnamefont {Celestre}}, \bibinfo {author}
  {\bibfnamefont {W.}~\bibnamefont {Chao}}, \bibinfo {author} {\bibfnamefont
  {K.}~\bibnamefont {Kaznatcheev}}, \bibinfo {author} {\bibfnamefont {K.~L.}\
  \bibnamefont {David}}, \bibinfo {author} {\bibfnamefont {F.}~\bibnamefont
  {Maia}}, \bibinfo {author} {\bibfnamefont {S.}~\bibnamefont {Marchesini}},
  \bibinfo {author} {\bibfnamefont {Y.~S.}\ \bibnamefont {Meng}}, \bibinfo
  {author} {\bibfnamefont {T.}~\bibnamefont {Warwick}}, \bibinfo {author}
  {\bibfnamefont {L.~L.}\ \bibnamefont {Yang}}, \ and\ \bibinfo {author}
  {\bibfnamefont {H.~A.}\ \bibnamefont {Padmore}},\ }\href {\doibase
  10.1038/nphoton.2014.207} {\bibfield  {journal} {\bibinfo  {journal} {Nat
  Photon}\ }\textbf {\bibinfo {volume} {8}},\ \bibinfo {pages} {765} (\bibinfo
  {year} {2014})}\BibitemShut {NoStop}%
\bibitem [{\citenamefont {Robinson}\ \emph {et~al.}(2001)\citenamefont
  {Robinson}, \citenamefont {Vartanyants}, \citenamefont {Williams},
  \citenamefont {Pfeifer},\ and\ \citenamefont {Pitney}}]{Robinson01}%
  \BibitemOpen
  \bibfield  {author} {\bibinfo {author} {\bibfnamefont {I.~K.}\ \bibnamefont
  {Robinson}}, \bibinfo {author} {\bibfnamefont {I.~A.}\ \bibnamefont
  {Vartanyants}}, \bibinfo {author} {\bibfnamefont {G.~J.}\ \bibnamefont
  {Williams}}, \bibinfo {author} {\bibfnamefont {M.~A.}\ \bibnamefont
  {Pfeifer}}, \ and\ \bibinfo {author} {\bibfnamefont {J.~A.}\ \bibnamefont
  {Pitney}},\ }\href {\doibase 10.1103/PhysRevLett.87.195505} {\bibfield
  {journal} {\bibinfo  {journal} {Physical Review Letters}\ }\textbf {\bibinfo
  {volume} {87}} (\bibinfo {year} {2001}),\
  10.1103/PhysRevLett.87.195505}\BibitemShut {NoStop}%
\bibitem [{\citenamefont {Thibault}\ \emph {et~al.}(2009)\citenamefont
  {Thibault}, \citenamefont {Dierolf}, \citenamefont {Bunk}, \citenamefont
  {Menzel},\ and\ \citenamefont {Pfeiffer}}]{Thib}%
  \BibitemOpen
  \bibfield  {author} {\bibinfo {author} {\bibfnamefont {P.}~\bibnamefont
  {Thibault}}, \bibinfo {author} {\bibfnamefont {M.}~\bibnamefont {Dierolf}},
  \bibinfo {author} {\bibfnamefont {O.}~\bibnamefont {Bunk}}, \bibinfo {author}
  {\bibfnamefont {A.}~\bibnamefont {Menzel}}, \ and\ \bibinfo {author}
  {\bibfnamefont {F.}~\bibnamefont {Pfeiffer}},\ }\href {\doibase
  10.1016/j.ultramic.2008.12.011} {\bibfield  {journal} {\bibinfo  {journal}
  {Ultramicroscopy}\ }\textbf {\bibinfo {volume} {109}},\ \bibinfo {pages}
  {338} (\bibinfo {year} {2009})}\BibitemShut {NoStop}%
\bibitem [{\citenamefont {Maiden}\ and\ \citenamefont
  {Rodenburg}(2009)}]{Rodenburg}%
  \BibitemOpen
  \bibfield  {author} {\bibinfo {author} {\bibfnamefont {A.~M.}\ \bibnamefont
  {Maiden}}\ and\ \bibinfo {author} {\bibfnamefont {J.~M.}\ \bibnamefont
  {Rodenburg}},\ }\href@noop {} {\bibfield  {journal} {\bibinfo  {journal}
  {Ultramicroscopy}\ }\textbf {\bibinfo {volume} {109}},\ \bibinfo {pages}
  {1256} (\bibinfo {year} {2009})}\BibitemShut {NoStop}%
\bibitem [{\citenamefont {Guizar-Sicairos}\ and\ \citenamefont
  {Fienup}(2008)}]{Guizar08}%
  \BibitemOpen
  \bibfield  {author} {\bibinfo {author} {\bibfnamefont {M.}~\bibnamefont
  {Guizar-Sicairos}}\ and\ \bibinfo {author} {\bibfnamefont {J.~R.}\
  \bibnamefont {Fienup}},\ }\href@noop {} {\bibfield  {journal} {\bibinfo
  {journal} {Optics Express}\ }\textbf {\bibinfo {volume} {16}},\ \bibinfo
  {pages} {7264} (\bibinfo {year} {2008})}\BibitemShut {NoStop}%
\bibitem [{\citenamefont {Shenfield}\ and\ \citenamefont
  {Rodenburg}(2011)}]{Shenfield11}%
  \BibitemOpen
  \bibfield  {author} {\bibinfo {author} {\bibfnamefont {A.}~\bibnamefont
  {Shenfield}}\ and\ \bibinfo {author} {\bibfnamefont {J.~M.}\ \bibnamefont
  {Rodenburg}},\ }\href {\doibase 10.1063/1.3600235} {\bibfield  {journal}
  {\bibinfo  {journal} {Journal of Applied Physics}\ }\textbf {\bibinfo
  {volume} {109}} (\bibinfo {year} {2011}),\ 10.1063/1.3600235}\BibitemShut
  {NoStop}%
\bibitem [{\citenamefont {Maiden}\ \emph
  {et~al.}(2012{\natexlab{a}})\citenamefont {Maiden}, \citenamefont {Humphry},
  \citenamefont {Sarahan}, \citenamefont {Kraus},\ and\ \citenamefont
  {Rodenburg}}]{Maiden12}%
  \BibitemOpen
  \bibfield  {author} {\bibinfo {author} {\bibfnamefont {A.~M.}\ \bibnamefont
  {Maiden}}, \bibinfo {author} {\bibfnamefont {M.~J.}\ \bibnamefont {Humphry}},
  \bibinfo {author} {\bibfnamefont {M.~C.}\ \bibnamefont {Sarahan}}, \bibinfo
  {author} {\bibfnamefont {B.}~\bibnamefont {Kraus}}, \ and\ \bibinfo {author}
  {\bibfnamefont {J.~M.}\ \bibnamefont {Rodenburg}},\ }\href@noop {} {\bibfield
   {journal} {\bibinfo  {journal} {Ultramicroscopy}\ }\textbf {\bibinfo
  {volume} {120}},\ \bibinfo {pages} {64} (\bibinfo {year}
  {2012}{\natexlab{a}})}\BibitemShut {NoStop}%
\bibitem [{\citenamefont {Zhang}\ \emph {et~al.}(2013)\citenamefont {Zhang},
  \citenamefont {Peterson}, \citenamefont {Vila-Comamala}, \citenamefont
  {Berenguer}, \citenamefont {Bean}, \citenamefont {Chen}, \citenamefont
  {Menzel}, \citenamefont {Robinson},\ and\ \citenamefont
  {Rodenburg}}]{Zhang13}%
  \BibitemOpen
  \bibfield  {author} {\bibinfo {author} {\bibfnamefont {F.}~\bibnamefont
  {Zhang}}, \bibinfo {author} {\bibfnamefont {I.}~\bibnamefont {Peterson}},
  \bibinfo {author} {\bibfnamefont {J.}~\bibnamefont {Vila-Comamala}}, \bibinfo
  {author} {\bibfnamefont {A.~D.~F.}\ \bibnamefont {Berenguer}}, \bibinfo
  {author} {\bibfnamefont {R.}~\bibnamefont {Bean}}, \bibinfo {author}
  {\bibfnamefont {B.}~\bibnamefont {Chen}}, \bibinfo {author} {\bibfnamefont
  {A.}~\bibnamefont {Menzel}}, \bibinfo {author} {\bibfnamefont {I.~K.}\
  \bibnamefont {Robinson}}, \ and\ \bibinfo {author} {\bibfnamefont {J.~M.}\
  \bibnamefont {Rodenburg}},\ }\href@noop {} {\bibfield  {journal} {\bibinfo
  {journal} {Optics Express}\ }\textbf {\bibinfo {volume} {21}},\ \bibinfo
  {pages} {13592} (\bibinfo {year} {2013})}\BibitemShut {NoStop}%
\bibitem [{\citenamefont {Thibault}\ and\ \citenamefont
  {Menzel}(2013)}]{Thibault13}%
  \BibitemOpen
  \bibfield  {author} {\bibinfo {author} {\bibfnamefont {P.}~\bibnamefont
  {Thibault}}\ and\ \bibinfo {author} {\bibfnamefont {A.}~\bibnamefont
  {Menzel}},\ }\href@noop {} {\bibfield  {journal} {\bibinfo  {journal}
  {Nature}\ }\textbf {\bibinfo {volume} {494}},\ \bibinfo {pages} {68}
  (\bibinfo {year} {2013})}\BibitemShut {NoStop}%
\bibitem [{\citenamefont {Hruszkewycz}\ \emph {et~al.}(2017)\citenamefont
  {Hruszkewycz}, \citenamefont {Allain}, \citenamefont {Holt}, \citenamefont
  {Murray}, \citenamefont {Holt}, \citenamefont {Fuoss},\ and\ \citenamefont
  {Chamard}}]{Hruszkewycz17}%
  \BibitemOpen
  \bibfield  {author} {\bibinfo {author} {\bibfnamefont {S.~O.}\ \bibnamefont
  {Hruszkewycz}}, \bibinfo {author} {\bibfnamefont {M.}~\bibnamefont {Allain}},
  \bibinfo {author} {\bibfnamefont {M.~V.}\ \bibnamefont {Holt}}, \bibinfo
  {author} {\bibfnamefont {C.~E.}\ \bibnamefont {Murray}}, \bibinfo {author}
  {\bibfnamefont {J.~R.}\ \bibnamefont {Holt}}, \bibinfo {author}
  {\bibfnamefont {P.~H.}\ \bibnamefont {Fuoss}}, \ and\ \bibinfo {author}
  {\bibfnamefont {V.}~\bibnamefont {Chamard}},\ }\href {\doibase
  10.1038/nmat4798} {\bibfield  {journal} {\bibinfo  {journal} {Nat Mater}\
  }\textbf {\bibinfo {volume} {16}},\ \bibinfo {pages} {244} (\bibinfo {year}
  {2017})}\BibitemShut {NoStop}%
\bibitem [{\citenamefont {Hruszkewycz}\ \emph {et~al.}(2016)\citenamefont
  {Hruszkewycz}, \citenamefont {Zhang}, \citenamefont {Holt}, \citenamefont
  {Highland}, \citenamefont {Evans},\ and\ \citenamefont
  {Fuoss}}]{Hruszkewycz16}%
  \BibitemOpen
  \bibfield  {author} {\bibinfo {author} {\bibfnamefont {S.~O.}\ \bibnamefont
  {Hruszkewycz}}, \bibinfo {author} {\bibfnamefont {Q.}~\bibnamefont {Zhang}},
  \bibinfo {author} {\bibfnamefont {M.~V.}\ \bibnamefont {Holt}}, \bibinfo
  {author} {\bibfnamefont {M.~J.}\ \bibnamefont {Highland}}, \bibinfo {author}
  {\bibfnamefont {P.~G.}\ \bibnamefont {Evans}}, \ and\ \bibinfo {author}
  {\bibfnamefont {P.~H.}\ \bibnamefont {Fuoss}},\ }\href {\doibase
  10.1103/PhysRevA.94.043803} {\bibfield  {journal} {\bibinfo  {journal} {Phys.
  Rev. A}\ }\textbf {\bibinfo {volume} {94}},\ \bibinfo {pages} {043803}
  (\bibinfo {year} {2016})}\BibitemShut {NoStop}%
\bibitem [{\citenamefont {Holt}\ \emph {et~al.}(2014)\citenamefont {Holt},
  \citenamefont {Hruszkewycz}, \citenamefont {Murray}, \citenamefont {Holt},
  \citenamefont {Paskiewicz},\ and\ \citenamefont {Fuoss}}]{Holt14}%
  \BibitemOpen
  \bibfield  {author} {\bibinfo {author} {\bibfnamefont {M.~V.}\ \bibnamefont
  {Holt}}, \bibinfo {author} {\bibfnamefont {S.~O.}\ \bibnamefont
  {Hruszkewycz}}, \bibinfo {author} {\bibfnamefont {C.~E.}\ \bibnamefont
  {Murray}}, \bibinfo {author} {\bibfnamefont {J.~R.}\ \bibnamefont {Holt}},
  \bibinfo {author} {\bibfnamefont {D.~M.}\ \bibnamefont {Paskiewicz}}, \ and\
  \bibinfo {author} {\bibfnamefont {P.~H.}\ \bibnamefont {Fuoss}},\ }\href@noop
  {} {\bibfield  {journal} {\bibinfo  {journal} {Physical review letters}\
  }\textbf {\bibinfo {volume} {112}},\ \bibinfo {pages} {165502} (\bibinfo
  {year} {2014})},\ \bibinfo {note} {0}\BibitemShut {NoStop}%
\bibitem [{\citenamefont {Hruszkewycz}\ \emph {et~al.}(2012)\citenamefont
  {Hruszkewycz}, \citenamefont {Holt}, \citenamefont {Murray}, \citenamefont
  {Bruley}, \citenamefont {Holt}, \citenamefont {Tripathi}, \citenamefont
  {Shpyrko}, \citenamefont {McNulty}, \citenamefont {Highland},\ and\
  \citenamefont {Fuoss}}]{Hruszkewycz12}%
  \BibitemOpen
  \bibfield  {author} {\bibinfo {author} {\bibfnamefont {S.~O.}\ \bibnamefont
  {Hruszkewycz}}, \bibinfo {author} {\bibfnamefont {M.~V.}\ \bibnamefont
  {Holt}}, \bibinfo {author} {\bibfnamefont {C.~E.}\ \bibnamefont {Murray}},
  \bibinfo {author} {\bibfnamefont {J.}~\bibnamefont {Bruley}}, \bibinfo
  {author} {\bibfnamefont {J.}~\bibnamefont {Holt}}, \bibinfo {author}
  {\bibfnamefont {A.}~\bibnamefont {Tripathi}}, \bibinfo {author}
  {\bibfnamefont {O.~G.}\ \bibnamefont {Shpyrko}}, \bibinfo {author}
  {\bibfnamefont {I.}~\bibnamefont {McNulty}}, \bibinfo {author} {\bibfnamefont
  {M.~J.}\ \bibnamefont {Highland}}, \ and\ \bibinfo {author} {\bibfnamefont
  {P.~H.}\ \bibnamefont {Fuoss}},\ }\href {\doibase 10.1021/nl303201w}
  {\bibfield  {journal} {\bibinfo  {journal} {Nano Letters}\ }\textbf {\bibinfo
  {volume} {12}},\ \bibinfo {pages} {5148} (\bibinfo {year}
  {2012})}\BibitemShut {NoStop}%
\bibitem [{\citenamefont {Hruszkewycz}\ \emph {et~al.}(2013)\citenamefont
  {Hruszkewycz}, \citenamefont {Highland}, \citenamefont {Holt}, \citenamefont
  {Kim}, \citenamefont {Folkman}, \citenamefont {Thompson}, \citenamefont
  {Tripathi}, \citenamefont {Stephenson}, \citenamefont {Hong},\ and\
  \citenamefont {Fuoss}}]{Hruszkewycz13}%
  \BibitemOpen
  \bibfield  {author} {\bibinfo {author} {\bibfnamefont {S.~O.}\ \bibnamefont
  {Hruszkewycz}}, \bibinfo {author} {\bibfnamefont {M.~J.}\ \bibnamefont
  {Highland}}, \bibinfo {author} {\bibfnamefont {M.~V.}\ \bibnamefont {Holt}},
  \bibinfo {author} {\bibfnamefont {D.}~\bibnamefont {Kim}}, \bibinfo {author}
  {\bibfnamefont {C.~M.}\ \bibnamefont {Folkman}}, \bibinfo {author}
  {\bibfnamefont {C.}~\bibnamefont {Thompson}}, \bibinfo {author}
  {\bibfnamefont {A.}~\bibnamefont {Tripathi}}, \bibinfo {author}
  {\bibfnamefont {G.~B.}\ \bibnamefont {Stephenson}}, \bibinfo {author}
  {\bibfnamefont {S.}~\bibnamefont {Hong}}, \ and\ \bibinfo {author}
  {\bibfnamefont {P.~H.}\ \bibnamefont {Fuoss}},\ }\href {\doibase
  10.1103/PhysRevLett.110.177601} {\bibfield  {journal} {\bibinfo  {journal}
  {Physical Review Letters}\ }\textbf {\bibinfo {volume} {110}} (\bibinfo
  {year} {2013}),\ 10.1103/PhysRevLett.110.177601}\BibitemShut {NoStop}%
\bibitem [{\citenamefont {Chamard}\ \emph {et~al.}(2015)\citenamefont
  {Chamard}, \citenamefont {Allain}, \citenamefont {Godard}, \citenamefont
  {Talneau}, \citenamefont {Patriarche},\ and\ \citenamefont
  {Burghammer}}]{ChamardSci2015}%
  \BibitemOpen
  \bibfield  {author} {\bibinfo {author} {\bibfnamefont {V.}~\bibnamefont
  {Chamard}}, \bibinfo {author} {\bibfnamefont {M.}~\bibnamefont {Allain}},
  \bibinfo {author} {\bibfnamefont {P.}~\bibnamefont {Godard}}, \bibinfo
  {author} {\bibfnamefont {A.}~\bibnamefont {Talneau}}, \bibinfo {author}
  {\bibfnamefont {G.}~\bibnamefont {Patriarche}}, \ and\ \bibinfo {author}
  {\bibfnamefont {M.}~\bibnamefont {Burghammer}},\ }\href@noop {} {\bibfield
  {journal} {\bibinfo  {journal} {Scientific Reports}\ }\textbf {\bibinfo
  {volume} {5}} (\bibinfo {year} {2015})}\BibitemShut {NoStop}%
\bibitem [{\citenamefont {Pateras}\ \emph {et~al.}(2015)\citenamefont
  {Pateras}, \citenamefont {Allain}, \citenamefont {Godard}, \citenamefont
  {Largeau}, \citenamefont {Patriarche}, \citenamefont {Talneau}, \citenamefont
  {Pantzas}, \citenamefont {Burghammer}, \citenamefont {Minkevich},\ and\
  \citenamefont {Chamard}}]{Pateras2015}%
  \BibitemOpen
  \bibfield  {author} {\bibinfo {author} {\bibfnamefont {A.~I.}\ \bibnamefont
  {Pateras}}, \bibinfo {author} {\bibfnamefont {M.}~\bibnamefont {Allain}},
  \bibinfo {author} {\bibfnamefont {P.}~\bibnamefont {Godard}}, \bibinfo
  {author} {\bibfnamefont {L.}~\bibnamefont {Largeau}}, \bibinfo {author}
  {\bibfnamefont {G.}~\bibnamefont {Patriarche}}, \bibinfo {author}
  {\bibfnamefont {A.}~\bibnamefont {Talneau}}, \bibinfo {author} {\bibfnamefont
  {K.}~\bibnamefont {Pantzas}}, \bibinfo {author} {\bibfnamefont
  {M.}~\bibnamefont {Burghammer}}, \bibinfo {author} {\bibfnamefont {A.~A.}\
  \bibnamefont {Minkevich}}, \ and\ \bibinfo {author} {\bibfnamefont
  {V.}~\bibnamefont {Chamard}},\ }\href@noop {} {\bibfield  {journal} {\bibinfo
   {journal} {Phys. Rev. B}\ }\textbf {\bibinfo {volume} {92}} (\bibinfo {year}
  {2015})}\BibitemShut {NoStop}%
\bibitem [{\citenamefont {Godard}\ \emph {et~al.}(2011)\citenamefont {Godard},
  \citenamefont {Carbone}, \citenamefont {Allain}, \citenamefont
  {Mastropietro}, \citenamefont {Chen}, \citenamefont {Capello}, \citenamefont
  {Diaz}, \citenamefont {Metzger}, \citenamefont {Stangl},\ and\ \citenamefont
  {Chamard}}]{Godard2011}%
  \BibitemOpen
  \bibfield  {author} {\bibinfo {author} {\bibfnamefont {P.}~\bibnamefont
  {Godard}}, \bibinfo {author} {\bibfnamefont {G.}~\bibnamefont {Carbone}},
  \bibinfo {author} {\bibfnamefont {M.}~\bibnamefont {Allain}}, \bibinfo
  {author} {\bibfnamefont {F.}~\bibnamefont {Mastropietro}}, \bibinfo {author}
  {\bibfnamefont {G.}~\bibnamefont {Chen}}, \bibinfo {author} {\bibfnamefont
  {L.}~\bibnamefont {Capello}}, \bibinfo {author} {\bibfnamefont
  {A.}~\bibnamefont {Diaz}}, \bibinfo {author} {\bibfnamefont {T.}~\bibnamefont
  {Metzger}}, \bibinfo {author} {\bibfnamefont {J.}~\bibnamefont {Stangl}}, \
  and\ \bibinfo {author} {\bibfnamefont {V.}~\bibnamefont {Chamard}},\
  }\href@noop {} {\bibfield  {journal} {\bibinfo  {journal} {Nature
  Communications}\ }\textbf {\bibinfo {volume} {2}} (\bibinfo {year}
  {2011})}\BibitemShut {NoStop}%
\bibitem [{\citenamefont {Stadler}\ \emph {et~al.}(2007)\citenamefont
  {Stadler}, \citenamefont {Harder}, \citenamefont {Robinson}, \citenamefont
  {Rentenberger}, \citenamefont {Karnthaler}, \citenamefont {Sepiol},\ and\
  \citenamefont {Vogl}}]{Stadler07}%
  \BibitemOpen
  \bibfield  {author} {\bibinfo {author} {\bibfnamefont {L.-M.}\ \bibnamefont
  {Stadler}}, \bibinfo {author} {\bibfnamefont {R.}~\bibnamefont {Harder}},
  \bibinfo {author} {\bibfnamefont {I.~K.}\ \bibnamefont {Robinson}}, \bibinfo
  {author} {\bibfnamefont {C.}~\bibnamefont {Rentenberger}}, \bibinfo {author}
  {\bibfnamefont {H.~P.}\ \bibnamefont {Karnthaler}}, \bibinfo {author}
  {\bibfnamefont {B.}~\bibnamefont {Sepiol}}, \ and\ \bibinfo {author}
  {\bibfnamefont {G.}~\bibnamefont {Vogl}},\ }\href {\doibase
  10.1103/PhysRevB.76.014204} {\bibfield  {journal} {\bibinfo  {journal}
  {Physical Review B}\ }\textbf {\bibinfo {volume} {76}} (\bibinfo {year}
  {2007}),\ 10.1103/PhysRevB.76.014204}\BibitemShut {NoStop}%
\bibitem [{\citenamefont {Bean}(2012)}]{BeanTh}%
  \BibitemOpen
  \bibfield  {author} {\bibinfo {author} {\bibfnamefont {R.}~\bibnamefont
  {Bean}},\ }\emph {\bibinfo {title} {Domain Structure Imaging with Bragg
  Geometry X-ray Ptychography}},\ \href@noop {} {Ph.D. thesis},\ \bibinfo
  {school} {University College of London} (\bibinfo {year} {2012})\BibitemShut
  {NoStop}%
\bibitem [{\citenamefont {Wildes}\ \emph {et~al.}(2001)\citenamefont {Wildes},
  \citenamefont {Mayer},\ and\ \citenamefont {Theis-Brohl}}]{Wildes01}%
  \BibitemOpen
  \bibfield  {author} {\bibinfo {author} {\bibfnamefont {A.~R.}\ \bibnamefont
  {Wildes}}, \bibinfo {author} {\bibfnamefont {J.}~\bibnamefont {Mayer}}, \
  and\ \bibinfo {author} {\bibfnamefont {K.}~\bibnamefont {Theis-Brohl}},\
  }\href {\doibase 10.1016/s0040-6090(01)01631-5} {\bibfield  {journal}
  {\bibinfo  {journal} {Thin Solid Films}\ }\textbf {\bibinfo {volume} {401}},\
  \bibinfo {pages} {7} (\bibinfo {year} {2001})}\BibitemShut {NoStop}%
\bibitem [{\citenamefont {Clavero}\ \emph {et~al.}(2012)\citenamefont
  {Clavero}, \citenamefont {Beringer}, \citenamefont {Roach}, \citenamefont
  {Skuza}, \citenamefont {Wong}, \citenamefont {Batchelor}, \citenamefont
  {Reece},\ and\ \citenamefont {Lukaszew}}]{Nb2012}%
  \BibitemOpen
  \bibfield  {author} {\bibinfo {author} {\bibfnamefont {C.}~\bibnamefont
  {Clavero}}, \bibinfo {author} {\bibfnamefont {D.~B.}\ \bibnamefont
  {Beringer}}, \bibinfo {author} {\bibfnamefont {W.~M.}\ \bibnamefont {Roach}},
  \bibinfo {author} {\bibfnamefont {J.~R.}\ \bibnamefont {Skuza}}, \bibinfo
  {author} {\bibfnamefont {K.~C.}\ \bibnamefont {Wong}}, \bibinfo {author}
  {\bibfnamefont {A.~D.}\ \bibnamefont {Batchelor}}, \bibinfo {author}
  {\bibfnamefont {C.~E.}\ \bibnamefont {Reece}}, \ and\ \bibinfo {author}
  {\bibfnamefont {R.~A.}\ \bibnamefont {Lukaszew}},\ }\href {\doibase
  10.1021/cg3001834} {\bibfield  {journal} {\bibinfo  {journal} {Crystal Growth
  \& Design}\ }\textbf {\bibinfo {volume} {12}},\ \bibinfo {pages} {2588}
  (\bibinfo {year} {2012})}\BibitemShut {NoStop}%
\bibitem [{\citenamefont {Fiory}\ \emph {et~al.}(1984)\citenamefont {Fiory},
  \citenamefont {C.}, \citenamefont {Feldman},\ and\ \citenamefont
  {K.}}]{IKR03}%
  \BibitemOpen
  \bibfield  {author} {\bibinfo {author} {\bibfnamefont {A.~T.}\ \bibnamefont
  {Fiory}}, \bibinfo {author} {\bibfnamefont {B.~J.}\ \bibnamefont {C.}},
  \bibinfo {author} {\bibfnamefont {L.~C.}\ \bibnamefont {Feldman}}, \ and\
  \bibinfo {author} {\bibfnamefont {R.~I.}\ \bibnamefont {K.}},\ }\href@noop {}
  {\bibfield  {journal} {\bibinfo  {journal} {J. Applied Physics}\ }\textbf
  {\bibinfo {volume} {56}} (\bibinfo {year} {1984})}\BibitemShut {NoStop}%
\bibitem [{\citenamefont {Barabash}\ \emph {et~al.}(2001)\citenamefont
  {Barabash}, \citenamefont {Donner},\ and\ \citenamefont
  {Dosch}}]{Barabash01}%
  \BibitemOpen
  \bibfield  {author} {\bibinfo {author} {\bibfnamefont {R.~I.}\ \bibnamefont
  {Barabash}}, \bibinfo {author} {\bibfnamefont {W.}~\bibnamefont {Donner}}, \
  and\ \bibinfo {author} {\bibfnamefont {H.}~\bibnamefont {Dosch}},\ }\href
  {\doibase 10.1063/1.1342215} {\bibfield  {journal} {\bibinfo  {journal}
  {Applied Physics Letters}\ }\textbf {\bibinfo {volume} {78}},\ \bibinfo
  {pages} {443} (\bibinfo {year} {2001})}\BibitemShut {NoStop}%
\bibitem [{\citenamefont {Flynn}(1988)}]{Flynn88}%
  \BibitemOpen
  \bibfield  {author} {\bibinfo {author} {\bibfnamefont {C.~P.}\ \bibnamefont
  {Flynn}},\ }\href {\doibase 10.1088/0305-4608/18/9/005} {\bibfield  {journal}
  {\bibinfo  {journal} {Journal of Physics F-Metal Physics}\ }\textbf {\bibinfo
  {volume} {18}},\ \bibinfo {pages} {L195} (\bibinfo {year}
  {1988})}\BibitemShut {NoStop}%
\bibitem [{\citenamefont {Shi}\ \emph {et~al.}(2012)\citenamefont {Shi},
  \citenamefont {Xiong}, \citenamefont {Huang}, \citenamefont {Harder},\ and\
  \citenamefont {Robinson}}]{Xiaowen2012}%
  \BibitemOpen
  \bibfield  {author} {\bibinfo {author} {\bibfnamefont {X.}~\bibnamefont
  {Shi}}, \bibinfo {author} {\bibfnamefont {G.}~\bibnamefont {Xiong}}, \bibinfo
  {author} {\bibfnamefont {X.}~\bibnamefont {Huang}}, \bibinfo {author}
  {\bibfnamefont {R.}~\bibnamefont {Harder}}, \ and\ \bibinfo {author}
  {\bibfnamefont {I.}~\bibnamefont {Robinson}},\ }\href
  {http://stacks.iop.org/1367-2630/14/i=6/a=063029} {\bibfield  {journal}
  {\bibinfo  {journal} {New Journal of Physics}\ }\textbf {\bibinfo {volume}
  {14}},\ \bibinfo {pages} {063029} (\bibinfo {year} {2012})}\BibitemShut
  {NoStop}%
\bibitem [{\citenamefont {Rodenburg}\ and\ \citenamefont
  {Faulkner}(2004)}]{Rodenburg04}%
  \BibitemOpen
  \bibfield  {author} {\bibinfo {author} {\bibfnamefont {J.~M.}\ \bibnamefont
  {Rodenburg}}\ and\ \bibinfo {author} {\bibfnamefont {H.~M.~L.}\ \bibnamefont
  {Faulkner}},\ }\href@noop {} {\bibfield  {journal} {\bibinfo  {journal}
  {Applied Physics Letters}\ }\textbf {\bibinfo {volume} {85}},\ \bibinfo
  {pages} {4795} (\bibinfo {year} {2004})}\BibitemShut {NoStop}%
\bibitem [{\citenamefont {Nashed}\ \emph {et~al.}(2014)\citenamefont {Nashed},
  \citenamefont {Vine}, \citenamefont {Peterka}, \citenamefont {Deng},
  \citenamefont {Ross},\ and\ \citenamefont {Jacobsen}}]{Nashed14}%
  \BibitemOpen
  \bibfield  {author} {\bibinfo {author} {\bibfnamefont {Y.~S.~G.}\
  \bibnamefont {Nashed}}, \bibinfo {author} {\bibfnamefont {D.~J.}\
  \bibnamefont {Vine}}, \bibinfo {author} {\bibfnamefont {T.}~\bibnamefont
  {Peterka}}, \bibinfo {author} {\bibfnamefont {J.}~\bibnamefont {Deng}},
  \bibinfo {author} {\bibfnamefont {R.}~\bibnamefont {Ross}}, \ and\ \bibinfo
  {author} {\bibfnamefont {C.}~\bibnamefont {Jacobsen}},\ }\href {\doibase
  10.1364/oe.22.032082} {\bibfield  {journal} {\bibinfo  {journal} {Optics
  Express}\ }\textbf {\bibinfo {volume} {22}},\ \bibinfo {pages} {32082}
  (\bibinfo {year} {2014})}\BibitemShut {NoStop}%
\bibitem [{\citenamefont {Guizar-Sicairos}\ \emph {et~al.}(2014)\citenamefont
  {Guizar-Sicairos}, \citenamefont {Johnson}, \citenamefont {Diaz},
  \citenamefont {Holler}, \citenamefont {Karvinen}, \citenamefont {Stadler},
  \citenamefont {Dinapoli}, \citenamefont {Bunk},\ and\ \citenamefont
  {Menzel}}]{Guizar14}%
  \BibitemOpen
  \bibfield  {author} {\bibinfo {author} {\bibfnamefont {M.}~\bibnamefont
  {Guizar-Sicairos}}, \bibinfo {author} {\bibfnamefont {I.}~\bibnamefont
  {Johnson}}, \bibinfo {author} {\bibfnamefont {A.}~\bibnamefont {Diaz}},
  \bibinfo {author} {\bibfnamefont {M.}~\bibnamefont {Holler}}, \bibinfo
  {author} {\bibfnamefont {P.}~\bibnamefont {Karvinen}}, \bibinfo {author}
  {\bibfnamefont {H.-C.}\ \bibnamefont {Stadler}}, \bibinfo {author}
  {\bibfnamefont {R.}~\bibnamefont {Dinapoli}}, \bibinfo {author}
  {\bibfnamefont {O.}~\bibnamefont {Bunk}}, \ and\ \bibinfo {author}
  {\bibfnamefont {A.}~\bibnamefont {Menzel}},\ }\href {\doibase
  10.1364/oe.22.014859} {\bibfield  {journal} {\bibinfo  {journal} {Optics
  Express}\ }\textbf {\bibinfo {volume} {22}},\ \bibinfo {pages} {14859}
  (\bibinfo {year} {2014})}\BibitemShut {NoStop}%
\bibitem [{\citenamefont {Odstrcil}\ \emph {et~al.}(2016)\citenamefont
  {Odstrcil}, \citenamefont {Baksh}, \citenamefont {Boden}, \citenamefont
  {Card}, \citenamefont {Chad}, \citenamefont {Frey},\ and\ \citenamefont
  {Brocklesby}}]{Odstrcil16}%
  \BibitemOpen
  \bibfield  {author} {\bibinfo {author} {\bibfnamefont {M.}~\bibnamefont
  {Odstrcil}}, \bibinfo {author} {\bibfnamefont {P.}~\bibnamefont {Baksh}},
  \bibinfo {author} {\bibfnamefont {S.~A.}\ \bibnamefont {Boden}}, \bibinfo
  {author} {\bibfnamefont {R.}~\bibnamefont {Card}}, \bibinfo {author}
  {\bibfnamefont {J.~E.}\ \bibnamefont {Chad}}, \bibinfo {author}
  {\bibfnamefont {J.~G.}\ \bibnamefont {Frey}}, \ and\ \bibinfo {author}
  {\bibfnamefont {W.~S.}\ \bibnamefont {Brocklesby}},\ }\href {\doibase
  10.1364/OE.24.008360} {\bibfield  {journal} {\bibinfo  {journal} {Opt.
  Express}\ }\textbf {\bibinfo {volume} {24}},\ \bibinfo {pages} {8360}
  (\bibinfo {year} {2016})}\BibitemShut {NoStop}%
\bibitem [{\citenamefont {Maiden}\ \emph
  {et~al.}(2012{\natexlab{b}})\citenamefont {Maiden}, \citenamefont {Humphry},
  \citenamefont {Sarahan}, \citenamefont {Kraus},\ and\ \citenamefont
  {Rodenburg}}]{MaidenA2012}%
  \BibitemOpen
  \bibfield  {author} {\bibinfo {author} {\bibfnamefont {A.}~\bibnamefont
  {Maiden}}, \bibinfo {author} {\bibfnamefont {M.}~\bibnamefont {Humphry}},
  \bibinfo {author} {\bibfnamefont {M.}~\bibnamefont {Sarahan}}, \bibinfo
  {author} {\bibfnamefont {B.}~\bibnamefont {Kraus}}, \ and\ \bibinfo {author}
  {\bibfnamefont {J.}~\bibnamefont {Rodenburg}},\ }\href {\doibase
  http://dx.doi.org/10.1016/j.ultramic.2012.06.001} {\bibfield  {journal}
  {\bibinfo  {journal} {Ultramicroscopy}\ }\textbf {\bibinfo {volume} {120}},\
  \bibinfo {pages} {64 } (\bibinfo {year} {2012}{\natexlab{b}})}\BibitemShut
  {NoStop}%
\bibitem [{\citenamefont {Suzuki}(2016)}]{BurdetSci2016}%
  \BibitemOpen
  \bibfield  {author} {\bibinfo {author} {\bibfnamefont {S.~K. H. M. B. N.
  T.~Y.}\ \bibnamefont {Suzuki}, \bibfnamefont {Akihiro}},\ }\href@noop {}
  {\bibfield  {journal} {\bibinfo  {journal} {Scientific Reports}\ }\textbf
  {\bibinfo {volume} {6}} (\bibinfo {year} {2016})}\BibitemShut {NoStop}%
\bibitem [{\citenamefont {Guizar-Sicairos}\ \emph {et~al.}(2011)\citenamefont
  {Guizar-Sicairos}, \citenamefont {Diaz}, \citenamefont {Menzel},\ and\
  \citenamefont {Bunk}}]{Guizar11}%
  \BibitemOpen
  \bibfield  {author} {\bibinfo {author} {\bibfnamefont {M.}~\bibnamefont
  {Guizar-Sicairos}}, \bibinfo {author} {\bibfnamefont {A.}~\bibnamefont
  {Diaz}}, \bibinfo {author} {\bibfnamefont {A.}~\bibnamefont {Menzel}}, \ and\
  \bibinfo {author} {\bibfnamefont {O.}~\bibnamefont {Bunk}},\ }\href {\doibase
  10.1117/12.903688} {\bibfield  {journal} {\bibinfo  {journal} {22nd Congress
  of the International Commission for Optics: Light for the Development of the
  World}\ }\textbf {\bibinfo {volume} {8011}} (\bibinfo {year} {2011}),\
  10.1117/12.903688}\BibitemShut {NoStop}%
\bibitem [{\citenamefont {Shapiro}\ \emph {et~al.}(2005)\citenamefont
  {Shapiro}, \citenamefont {Thibault}, \citenamefont {Beetz}, \citenamefont
  {Elser}, \citenamefont {Howells}, \citenamefont {Jacobsen}, \citenamefont
  {Kirz}, \citenamefont {Lima}, \citenamefont {Miao}, \citenamefont {Neiman},\
  and\ \citenamefont {Sayre}}]{Shapiro05}%
  \BibitemOpen
  \bibfield  {author} {\bibinfo {author} {\bibfnamefont {D.}~\bibnamefont
  {Shapiro}}, \bibinfo {author} {\bibfnamefont {P.}~\bibnamefont {Thibault}},
  \bibinfo {author} {\bibfnamefont {T.}~\bibnamefont {Beetz}}, \bibinfo
  {author} {\bibfnamefont {V.}~\bibnamefont {Elser}}, \bibinfo {author}
  {\bibfnamefont {M.}~\bibnamefont {Howells}}, \bibinfo {author} {\bibfnamefont
  {C.}~\bibnamefont {Jacobsen}}, \bibinfo {author} {\bibfnamefont
  {J.}~\bibnamefont {Kirz}}, \bibinfo {author} {\bibfnamefont {E.}~\bibnamefont
  {Lima}}, \bibinfo {author} {\bibfnamefont {H.}~\bibnamefont {Miao}}, \bibinfo
  {author} {\bibfnamefont {A.~M.}\ \bibnamefont {Neiman}}, \ and\ \bibinfo
  {author} {\bibfnamefont {D.}~\bibnamefont {Sayre}},\ }\href {\doibase
  10.1073/pnas.0503305102} {\bibfield  {journal} {\bibinfo  {journal}
  {Proceedings of the National Academy of Sciences}\ }\textbf {\bibinfo
  {volume} {102}},\ \bibinfo {pages} {15343} (\bibinfo {year} {2005})},\
  \Eprint
  {http://arxiv.org/abs/http://www.pnas.org/content/102/43/15343.full.pdf}
  {http://www.pnas.org/content/102/43/15343.full.pdf} \BibitemShut {NoStop}%
\bibitem [{\citenamefont {Manuel Guizar-Sicairos}\ and\ \citenamefont
  {Fienup}(2008)}]{subpixel}%
  \BibitemOpen
  \bibfield  {author} {\bibinfo {author} {\bibfnamefont {S.~T.~T.}\
  \bibnamefont {Manuel Guizar-Sicairos}}\ and\ \bibinfo {author} {\bibfnamefont
  {J.~R.}\ \bibnamefont {Fienup}},\ }\href@noop {} {\bibfield  {journal}
  {\bibinfo  {journal} {Opt. Lett.}\ }\textbf {\bibinfo {volume} {33}},\
  \bibinfo {pages} {156} (\bibinfo {year} {2008})}\BibitemShut {NoStop}%
\end{thebibliography}%

%ADD:

\end{document}